\documentstyle[12pt]{article}
\begin{document}
\begin{center}
{\bf Mixed-spin systems: coexistence of Haldane gap\\
 and antiferromagnetic long range order\\}
Indrani Bose and Emily Chattopadhyay\\
Department of Physics\\
Bose Institute\\
93/1, Acharya Prafulla Chandra Road\\
Calcutta - 700009, India

\end{center}

\begin{abstract}

Recent experiments on the quasi-1D antiferromagnets $R_{2}BaNiO_{5}$ ( R = rare 
earth )  have shown the existence of purely 1D Haldane gap excitations propagating on 
the Ni chains. Below an ordering temperature,  the gap excitations survive and coexist
with the conventional spin waves in the ordered phase. We construct a model mixed-spin
  system in 2D for which the ground state can be exactly specified. Using the Matrix
Product Method, we show the existence of Haldane gap excitations in the ordered phase. 
 We consider different cases of ordering to study the effect of ordering on the degeneracy
  of the Haldane gap excitations.

\end{abstract}
P.A.C.S. Nos: 75.10 Jm, 75.50 Ee
\newpage
\section*{I. Introduction}

Haldane$^1$ in 1983 predicted that a Heisenberg  antiferromagnetic (AFM) chain of integer 
spins has a gap in the excitation spectrum. Later, Haldane's conjecture was verified in 
several theoretical as well as experimental studies.$^{2-4}$  In a quasi-1D magnetic system, the
 integer spin chains are coupled via weak exchange interactions. If the strength of the 
coupling is very small,  there is no possibility of ordering and individual spin chains
remain in the disordered Haldane gap phase at all temperatures. In materials like $CsNiCl_{3}$, the 
inter-chain coupling is sufficiently strong to give rise to an AFM ordered phase below
the N\'{e}el temperature $T_{N}.^{5-6}$  For temperature $T > T_{N}$, the three Haldane gap 
modes can be observed in experiments but as T approaches $T_{N}$, the gap vanishes at the
3D magnetic zone centre. In the ordered phase, two of the three modes become conventional 
gapless spin waves, while the third `longitudinal' mode develops a gap.

 For  quasi-1D Haldane gap systems, there now exists a third possibility. Recent inelastic 
neutron-scattering experiments$^{7-8}$ on the mixed-spin antiferromagnets $R_2BaNiO_5$ 
($R$ = $Pr$, $Nd$ or 
$Nd_xY_{1-x}$), provide clear evidence that the Haldane gap excitations propagating on the 
Ni chains survive even in the ordered phase. Unlike in systems like $CsNiCl_3$, the Haldane
gap does not become zero as T approaches $T_N$. In the ordered phase, the gapped Haldane
 modes coexist with the conventional spin waves characteristic of the ordered phase.The
 Haldane gap excitations maintain their 1D character in the entire phase diagram. In this paper, 
we construct a model mixed spin system in 2D for which the ground state can be exactly 
specified. AFM long-range order exists in the ground state with gapless spin waves as 
excitations. The system also consists of integer spin chains, the ground states of which are 
the Valence Bond Solid (VBS) states. Haldane gap excitations can be created in the chains in
the ordered phase of the mixed-spin system. We also discuss how the gap changes as the 
nature of the ordering changes.

\section*{II. Ground state and excitations}

Our model mixed spin system has an underlying lattice structure shown in Fig. 1. It basically
consists of a square lattice at each site of which a spin-1/2 (A-spin, solid circle) is sitting. The
centre of each square plaquette is occupied by a S=1 spin (B-spin, solid square). The B-spins
along a chain in the horizontal direction interact via nearest-neighbour (n.n.) interactions but the 
individual B-spin chains do not interact with each other. The B-spins interact with the n.n. A-spins
 along the diagonal lines in Fig. 1. The A-spins interact with n.n.  A-spins  along the horizontal
and vertical bonds of the square lattice. In $R_2BaNiO_5$ also, the S=1 Ni chains are 
decoupled and the $R^{+}_{3}$ sites with half-odd integer spins are positioned between the 
chains. The $R^{+}_{3}$ spins form a network via Ni-O-R and R-O-R superexchange routes.

We now demand that the B-spin chains are in Valence Bond Solid (VBS) ground states.
Affleck, Kenedy, Lieb and Tasaki (AKLT)$^4$ constructed a Hamiltonian for which the VBS state
is the exact ground state of an integer spin chain. We follow the same principle to construct
the Hamiltonian of our model. Fig. 2 shows two successive rows of A and B-spins. Consider 
the B-spins to be in a VBS state. Each spin-1 can be considered to be a symmetric 
combination of two spin-1/2's. In the VBS state, each spin-1/2 forms a singlet with a spin-1/2
at a neighbouring site. This process has to be symmetrized so that a spin-1 is restored at 
each site. In the ground state, a valence bond (VB) or singlet covers every link along the 
integer-spin chain so that the total spin of the state is zero. The Hamiltonian for which the 
VBS state is the exact ground state is given by
\begin{equation}
H_{AB}^{i} = \sum_{\Delta }^{}{P_{5/2}(\vec{S_a}+\vec{S_b}+\vec{S_c})}
\end{equation}
$P_{5/2}$ is the projection operator onto spin 5/2 for the triangle of spins. The sum is 
over successive triangles of spins with bases along the integer spin chain. The spins 
at sites $B_1$ and $B_2$ have magnitude 1 and the A-spin  has magnitude 1/2. In the
VBS state, the sum of the three spins can never be 5/2 as a VB exists between the basal 
spins. Thus, the projection operator $P_{5/2}$ operating on the triangle of spins in the VBS 
state gives zero. The VBS state is the exact ground state of $H_{AB}^i$ with eigenvalue zero.
 It can be easily verified that the A-spins  are free when the Hamiltonian in (1)
operates on the VBS state of the integer spin chain. The Hamiltonian $H_{AB}^i$, on expanding,
 is given by

\begin{equation}
H_{AB}^{i} = \frac{1}{20}  [ 2 + 5 ( \vec{S_a}.\vec{S_b} + \vec{S_b}.\vec{S_c} + \vec{S_c}.\vec{S_a} )
                              + 2 ( \vec{S_a}.\vec{S_b} + \vec{S_b}.\vec{S_c} + \vec{S_c}.\vec{S_a} )^2 ]
\end{equation}
When we consider the full 2D spin system, the Hamiltonian is given by
\[
H_{AB} = \sum_{i}^{}{H_{AB}^i}
\] The sum is  over individual B-spin chains.
 We now consider the interactions between the A-spins. The Hamiltonian $H_A$ is given by the 
 usual Heisenberg AFM Hamiltonian defined on the square lattice:

\begin{equation}
H_A = J \sum_{<ij>}^{}{\vec{S_i}.\vec{S_j}}
\end{equation} 
The spins have magnitude 1/2 and $<ij>$  are n.n.s. The ground state of (3), though not exactly 
known, has long range N\'{e}el type of order. The total Hamiltonian H is

\begin{equation}
H = H_A + H_{AB}
\end{equation}
The ground state of H has the following structure:  the  B-spin chains are in VBS states.
 The square network of A-spins is in the state which is the ground state of the S=1/2, n.n. 
Heisenberg AFM Hamiltonian on a square lattice. The decoupling of the S=1 spin and the S=1/2
spin ground states is possible because the A-spins are free when $H_{AB}$ operates on
the VBS states.

We now consider the excitations in the system. Spin wave excitations created in the A
network of spins have no effect on the VBS ground states of the chains of B-spins.
The spin wave excitations are gapless. Consider now the case when the network of A-spins
is in the ground state and an excitation is created in a B-spin chain. Again, for convenience,
we consider the two-chain strip of Fig. 2. The A-spins in the top chain are in the N\'{e}el state
$\mid \uparrow \downarrow \uparrow \downarrow \uparrow \downarrow \cdots \rangle$ indicative
of long range AFM order. We determine the excitation spectrum by the method of Matrix
Products$^{9-11}$ The VBS ground state of the chain of B-spins can be written as

\begin{equation}
\mid \psi _G\rangle  =  Tr ( g_1\otimes g_2\otimes \cdots \otimes g_L )
\end{equation}
where L is the number of sites (spins) in the chain. The $2\times 2$ matrix $g_i$ is given by

\begin{equation}
g_i  =  \left(\begin{array}{cc}
-\mid 0\rangle & -\sqrt{2}\mid 1\rangle \\
 \sqrt{2}\mid \bar{1}\rangle&  \mid 0\rangle 
\end{array}\right)
\end{equation}
The states $\mid 1\rangle,  \mid 0\rangle,  \mid \bar{1}\rangle$ correspond to $S^z$ = 1, 0
 and $-1$ respectively.
One can easily verify that $H_{AB}\mid \psi _G\rangle = 0$.  An excited state of the chain of B-spins
is constructed as [3]
\begin{equation}
\mid \psi _{B}^{a}(k)\rangle = \frac{1}{\sqrt{L}} \sum_{j=1}^{L}{e^{-ikj}\mid \psi _{Bj}^a\rangle} 
\end{equation}
$\mid \psi _{Bj}^a\rangle$ is obtained by replacing $g_j$ in the matrix product of the VBS state
by the matrix $g_j^a$,  where a = 1, 0, $ -1$ are the values of the z-component of the total spin,
$S^z$, of the excited state. The matrices $g_j^a$ have the form

\begin{equation}
g^1 = \left(\begin{array}{cc}
\sqrt{2} \mid 1\rangle&   0 \\
-\mid 0\rangle &   0 
\end{array}\right),
g^0 = \left(\begin{array}{cc}
-\mid 0\rangle &\sqrt{2}\mid 1\rangle  \\
 \sqrt{2}\mid \bar{1}\rangle& -\mid 0\rangle 
\end{array}\right),
g^{-1} = \left(\begin{array}{cc}
0     & -\mid 0\rangle \\
0     &\sqrt{2}\mid \bar{1}\rangle
\end{array}\right)
\end{equation}

\begin{equation}
\mid \psi _{Bj}^a\rangle = Tr g_1\otimes g_2\otimes \cdots \otimes g_{j -1}\otimes g^a_j\otimes g_{j+1}
\otimes \cdots \otimes g_L
\end{equation}

\begin{equation}
\left\langle\psi _B^a(k)\mid \psi _B^a(k)\right\rangle = \frac{1}{L} \sum_{l=1}^{L}{\sum_{p=1}^{L}
{e^{ik(l-p)}\left\langle\psi _{Bl}^a\mid \psi _{Bp}^a\right\rangle}}
\end{equation}
We now consider the case a = 1.
\begin{eqnarray}
\left\langle\psi _{Bl}^1\mid \psi _{Bp}^1\right\rangle  = \sum_{\left\{n_\alpha ,m_\alpha \right\}}^
{}{g_{n_1n_2}^{\dag}g_{n_2n_3}^{\dag}\cdots g_{n_ln_{l+1}}^{1\dagger} g_{n_{l+1}n_{l+2}}^\dagger 
\cdots g_{n_Ln_1}^\dagger}\nonumber \\g_{m_1m_2}g_{m_2m_3}\cdots g^1_{m_pm_{p+1}}\cdots 
g_{m_Lm_1}
\end{eqnarray} 
Define three $4\times 4$ matrices G, $G^{'}$and $G^{''}$ as
\begin{eqnarray}
G_{\mu _1\mu _2}& = &G_{(n_1m_1)(n_2m_2)} = g_{n_1n_2}^\dagger g_{m_1m_2}\nonumber\\
G^{'}_{\mu _1\mu _2}& =& G^{'}_{(n_lm_l)(n_{l+1}m_{l+1})} = g_{n_ln_{l+1}}^{1\dagger} 
g_{m_lm_{l+1}}\\
G^{''}_{\mu _1\mu _2} &=& G^{''}_{(n_pm_p)(n_{p+1}m_{p+1})} = g^\dagger _{n_pn_{p+1}}
g^{1}_{m_pm_{p+1}}\nonumber
\end{eqnarray}
\newpage
The ordering of multi-indices is given by

\begin{equation}
\mu  = 1,2,3,4 \longleftrightarrow  (11),(12),(21),(22)
\end{equation}

\begin{eqnarray}
\left\langle\psi _{Bl}^1(k)\mid \psi _{Bp}^1(k)\right\rangle & = &\sum_{\left\{n_\alpha ,m_\alpha 
\right\}}^{}{G_{(n_1m_1)(n_2m_2)}G_{(n_2m_2)(n_3m_3)}}\cdots \nonumber \\ 
&\cdots & G^{'}_{(n_lm_l)(n_{l+1}m_{l+1})}G_{(n_{l+1}m_{l+1})(n_{l+2}m_{l+2})}\cdots\nonumber \\
&\cdots & G^{''}_{(n_pm_p)(n_{p+1}m_{p+1})}\cdots 
G_{(n_Lm_L)(n_1m_1)}\nonumber\\ & & \nonumber\\
&= &TrG^{l-1}G^{'}G^{p-l+1}G^{''}G^{L-p} 
\end{eqnarray}
The matrices G, $G^{'}$ and $G^{''}$ are
\begin{eqnarray}
G&= &\left(\begin{array}{rrrr}
1  &0  &0  &2  \\
0  &-1  &0  &0  \\
0  &0  &-1  &0  \\
2  &0  &0  &1 
\end{array}\right) \nonumber  \\
 G^{'}& = & \left(\begin{array}{rrrr}
0 &-2  &0  &0  \\
0  &0  &0  &0  \\
1  &0  &0  &0  \\
0 &-1  &0  &0 
\end{array}\right)  \\
 G^{''}& = & \left(\begin{array}{rrrr}
 0  &0  &-2  &0  \\
1   &0   &0  & 0 \\
0   &0   &0  &0  \\
0   &0  &-1  &0 
\end{array}\right) \nonumber
\end{eqnarray}
The eigenvalues and eigenvectors of G are :
\begin{eqnarray}
\lambda _1 = 3,   &\lambda _2 = \lambda _3 = \lambda _4 = -1  &  \nonumber\\
\mid e_1\rangle  =\frac{1}{\sqrt{2}}\left(\begin{array}{l}
1 \\
0 \\
0 \\
1
\end{array}\right) &\mid e_2\rangle = \frac{1}{\sqrt{2}}\left(\begin{array}{l}
-1 \\
0 \\
0 \\
1
\end{array}\right)&    \\
\mid e_3\rangle =\left(\begin{array}{l}
0 \\
1 \\
0 \\
0
\end{array}\right)  &\mid e_4\rangle = \left(\begin{array}{l}
0 \\
0 \\
1 \\
0
\end{array}\right)&   \nonumber
\end{eqnarray}
\newpage
From (14) and in the thermodynamic limit L$\to \infty $,
\begin{eqnarray}
\left\langle\psi _{Bl}^1(k)\mid \psi _{Bp}^1\right\rangle   &=& \sum_{n=1 }^{4}{\lambda _1^{l-1}\left\langle e_1\left| G^{'}\right|n\right\rangle} \lambda
 _n^{p-l+1}\left\langle n \left| G^{''} \right|e_1\right\rangle \lambda _1^{L-p}\nonumber   \\
& &\nonumber \\
& = &\frac{1}{2}3^L(-\frac{1}{3})^{p-l}\nonumber 
\end{eqnarray}
 Therefore,

\begin{eqnarray}
\left\langle\psi _B^1(k)\mid \psi _B^1(k)\right\rangle 
&  =& \frac{1}{L}\sum_{l=1}^{L}{\sum_{p=1}^{L}{e^{ik(l-p)}\frac{1}{2}3^L
(-\frac{1}{3})^{\left|p-l\right|}}}\nonumber  \\
&=& 3^L\frac{2}{5+3\cos k}  
\end{eqnarray}
The chain of A-spins is in the N\'eel state $\mid \psi _{N\acute{e}el}\rangle$. The total wave function $\mid \psi \rangle$ of
the two-chain system is a product of $\mid \psi _B^1(k)\rangle$ and $\mid \psi _{N\acute{e}el}
\rangle$.The norm $\left\langle\psi \mid \psi \right\rangle$ is given by the same expression 
as in (17).
The excitation energy measured w.r.t.  the ground state energy is 
\begin{equation}
\omega _1(k) = \frac{\left\langle\psi \left|H_{AB}\right|\psi \right\rangle}{\left\langle\psi 
\mid \psi \right\rangle}
\end{equation}
The denominator has already been calculated. We now calculate the numerator. $H_{AB}$ for
single chain is given by Eq. (2). The two nearest-neighbour sites of the lower chain in 
Fig. 2 and the intermediate site in the upper chain form the vertices of a triangle. If one of the
sites of the lower chain has the `defect' matrix $g^{1}$ assosiated with it, then there are four
possible states of the triangle :
\begin{eqnarray}
\mid \varphi _{1j}\rangle\equiv \mid\begin{array}{c}
\uparrow  \\
g_j^{1}\otimes g_{j+1}
\end{array}\rangle, &{\large \mid}\varphi _{2j}\rangle \equiv \mid\begin{array}{c}
\uparrow  \\
g_j\otimes g_{j+1}^{1}
\end{array} \rangle & \nonumber \\
 \mid \varphi _{3j}\rangle \equiv \mid\begin{array}{c}
\downarrow  \\
g^{1}_j\otimes g_{j+1}
\end{array}\rangle,& \mid \varphi _{4j}\rangle \equiv \mid\begin{array}{c}
\downarrow  \\
g_j\otimes g_{j+1}^{1}
\end{array}\rangle & 
\end{eqnarray}
One can verify that

\begin{eqnarray*}
H_{j,j+1}\mid \varphi _{2j}\rangle &=& 0 \\
H_{j,j+1}\mid\varphi_{4j}\rangle &=& 0
\end{eqnarray*}
\begin{eqnarray}
H_{j,j+1}\mid \varphi _{1j}\rangle &=& \uparrow \frac{1}{20}\left(\begin{array}{cc}
-8\sqrt{2}\left(\mid 10\rangle +\mid 01\rangle\right) & -40\mid 11\rangle  \\
 8\mid 00\rangle + 4\mid \bar{1}1\rangle + 4\mid 1\bar{1}\rangle&  8\sqrt{2}\left(\mid 
01\rangle + \mid 10\rangle\right)
\end{array}\right)   \nonumber \\ &&\nonumber\\
&&+ \downarrow \frac{1}{20} \left(\begin{array}{cc}
-8\mid 11\rangle & 0 \\
4\sqrt{2}\left(\mid 01\rangle + \mid 10\rangle\right) & 8\mid 11\rangle
\end{array}\right)  \nonumber\\ &&\nonumber\\
H_{j,j+1}\mid \varphi _{3j}\rangle &=&  \uparrow \frac{1}{20} \left(\begin{array}{cc}
-8\mid 00\rangle -4\mid 1\bar{1} \rangle - 4\mid \bar{1}1\rangle &  -8\sqrt{2}\left(\mid 10\rangle
 + \mid 01\rangle\right)\\
 4\sqrt{2}\left(\mid \bar{1}0\rangle + \mid 0\bar{1}\rangle\right)& 8\mid 00\rangle + 4\mid 1
\bar{1}\rangle + 4\mid \bar{1}1\rangle 
\end{array}\right)  \nonumber\\ &&\nonumber\\
&&+  \downarrow \frac{1}{20}\left(\begin{array}{cc}
-4\sqrt{2}\left(\mid 10\rangle + \mid 01\rangle\right) & -8\mid 11\rangle \\
8\mid 00\rangle + 4\mid 1\bar{1}\rangle + 4 \mid \bar{1}1\rangle & 4\sqrt{2}\left(\mid 01\rangle +
\mid 10\rangle\right)
\end{array}\right)   
\end{eqnarray}
The up and down spins outside the brackets represent the states of the A-spin.

We represent $\mid \psi \rangle$ by the state
\begin{equation}
\mid \psi \rangle = \mid \begin{array}{c}
\psi _{N\acute{e}el} \\
\psi _B^1(k)
\end{array}\rangle
\end{equation}
in which the upper and lower rows contain the states of the upper and lower
 chains respectively. One can verify that

\begin{eqnarray}
& &\left\langle\begin{array}{c}
\psi _{N\acute{e}el} \\
\psi ^{1}_{Bj^{'}}(k)
\end{array}\left|H_{AB}\right|\begin{array}{c}
\psi _{N\acute{e}el} \\
\psi ^{1}_{Bj}(k)
\end{array}\right\rangle  \nonumber \\ &&\nonumber\\
&=& \delta _{jj^{'}}\left\langle\begin{array}{c}
\psi _{N\acute{e}el} \\
\psi ^1_{Bj}(k)
\end{array}\left|\left(H_{AB}\right)_{j j+1}\right|\begin{array}{c}
\psi _{N\acute{e}el} \\
\psi ^1_{Bj}(k)
\end{array}\right\rangle \nonumber  \\ &&\nonumber\\
&=& \delta _{jj^{'}} Tr G^{j-1}Z\left(H\right)G^{L-j-1} 
\end{eqnarray}
where the matrix G is given in (15) and 
\begin{eqnarray}
Z\left(H\right)_{\mu _1\mu _2} = Z\left(H\right)_{(n_1m_1)(n_2m_2)}  & & \nonumber \\
= (\varphi _{1j}^\dagger )_{n_1n_2}[(H_{AB})_{j j+1}\varphi _{1j}]_{m_1m_2} & (j  \,\,odd) &\nonumber \\
= (\varphi ^\dagger _{3j})_{n_1n_2}[(H_{AB})_{j j+1}\varphi _{3j}]_{m_1m_2} & (j \,\,   even) & 
\end{eqnarray}
From Eqns. (20)
\begin{eqnarray}
 Z^{odd}(H)& =&\frac{1}{20}\left(\begin{array}{cccc}
16 &0  &0  &80  \\
0 &-16  &0  &0  \\
0 &0  &-16  &0  \\
8 &0  &0  &16 
\end{array}\right)\nonumber  \\
Z^{even}(H)& =& \frac{1}{20}\left(\begin{array}{cccc}
8 &0  &0  &16  \\
0 &-8  &0  &0  \\
0 &0  & -8 &0  \\
8 &0  &0  &8 
\end{array}\right)  
\end{eqnarray}
In the limit $L\to \infty$ , Eq. (22) reduces to
\begin{eqnarray}
& &\left\langle\begin{array}{c}
\psi _{N\acute{e}el} \\
\psi ^1_{Bj}(k)
\end{array}\left|(H_{AB})_{jj+1}\right|\begin{array}{c}
\psi _{N\acute{e}el} \\
\psi ^1_{Bj}(k)
\end{array}\right\rangle \nonumber  \\ &&\nonumber\\
& =& \left\langle e_1 \left|G^{L-2} Z(H)\right| e_1\right\rangle\nonumber   \\ &&\nonumber\\
&=& \lambda _1^{L-2} \left\langle e_1\left|Z(H)\right|e_1\right\rangle  \nonumber\\ &&\nonumber\\
&=& \frac{1}{20}\times 3^{L-2}\times 60\quad for \quad j = odd\nonumber\\
&=&\frac{1}{20}\times 3^{L-2}\times 20\quad for \quad j = even
\end{eqnarray}
$\lambda _1$ and $\mid e_1\rangle$ are given in Eq. (16). From (7),
\begin{eqnarray}
&  &\left\langle\psi \left|H_{AB}\right|\psi \right\rangle\nonumber\\
&=& \frac{1}{L}\sum_{j^{'}}^{}{\sum_{j}^{}{e^{ik(j^{'}-j)}\delta _{jj^{'}}\left\langle\begin{array}{c}
\psi _{N\acute{e}el}  \\
\psi ^1_{Bj}(k)
\end{array}\left|(H_{AB})_{jj+1}\right|\begin{array}{c}
\psi _{N\acute{e}el} \\
\psi ^1_{Bj}(k)
\end{array}\right\rangle}}\nonumber\\ &&\nonumber\\
& =& \frac{1}{20}\times \frac{3^{L-2}}{L}(\frac{L}{2}60 + \frac{L}{2}20)\nonumber \\ &&\nonumber\\
&= &3^{L-2}\times 2 
\end{eqnarray}
From (18), (17) and (26), one gets
\begin{equation}
\omega _1(k) = \frac{1}{9}(5 + 3 \cos k )
\end{equation}
The excitation energy (27) has been determined for a two-chain strip. 
For the 2D spin system, the chain of B-spins is connected to two chains of A-spins and the
excitation energy is twice of that in (27). The excitation energy $\omega _{-1}(k)$ for $S^z
= -1$ is the same as $\omega _1(k)$. The excitation energy $\omega _0(k)$ for $S^z = 0$
is  $2\omega _1(k)$.  Thus the gaps in the excitation spectrum for the 2D spin system are :
\begin{equation}
\Delta _{\pm 1} = \frac{4}{9}\quad , \quad \Delta _0 = \frac{8}{9}
\end{equation}

We now consider the case when excitations are created in both the A and B-chains. The 
excitation in the A-chain travels in the network of A-spins as a 2D spin wave. The excitation
spectrum of the spin wave is gapless. The Haldane gap excitation is confined to the B-spin chain.
For a ferromagnetic ordering of A-spins, one can verify that  $\Delta _{+1}\neq \Delta _{-1}$,
i.e., the degeneracy of the triplet Haldane modes is fully lifted.
\section*{III. Conclusion}

We have constructed a model mixed spin system for which the ground state can be exactly
 specified. The coexistence of AFM long range order and Haldane gap excitations has been 
explicitly shown. In the ordered phase, the degeneracy of the three Haldane gap modes is partially lifted. 
The transverse modes $(S^z=\pm 1)$  are still degenerate whereas the longitudinal mode 
$(S^z=0 )$ has an excitation gap twice that of the transverse modes. These results are
in agreement with the experimental observations  on the $R_2BaNiO_5$ systems.$^{7-8}$
In these systems, coexistence of gapless spin waves in an ordered phase and the Haldane 
gap excitations confined to Ni chains, has been observed. The three Haldane gap modes
in an isolated chain of integer spins are degenerate. The lifting of the degeneracy seen in
 $R_2BaNiO_5$   is due to the ordering of the rare-earth spins.$^8$ It has been suggested$^{12}$
that in $R_2BaNiO_5$ the Ni-spins participate in the long range order. This is, however, not
so in our model, as in the ground state there is a complete decoupling of the A and B-spins.
 An important feature of the $R_2BaNiO_5$ systems is the increase in the gap energy
as the temperature is lowered, i.e., as the system becomes more ordered$^{12}$. Finite 
temperature calculations on our model are in progress and the results will be 
reported elsewhere.
\section*{Acknowledgement}

One of the Authors (EC) is supported by the Council of Scientific and Industrial Research,
India under sanction No. 9/15(186)/97-EMR-I.
\newpage
\section*{References}

$^1$F.D.M. Haldane, Phys. Lett. 93 A,  464  (1983), Phys. Rev. Lett. 50, 1153 

(1983)\\
$^2$L. P. Regnault, I.  Zaliznyak, J. P. Renard and C. Vettier, Phys. Rev. B 50,

 9174 (1994)
and references therein\\
$^3$G. F\'ath and J. S\'olyom, J. Phys.: Condens. Matter 5, 8983 (1993) and 

references
therein\\
$^4$I. Affleck, T. Kennedy, E. H. Lieb and H. Tasaki, Phys. Rev. Lett. 59,  799  

(1987);
Commun. Math. Phys. 115, 477 (1988)\\
$^5$W. J. L.  Buyers et al., Phys. Rev. Lett. 56, 371 (1986); R. M. Morra,  

W. J. L Buyers, R. L.
Armstrong and K. Hirakawa, Phys. Rev. B 38, 

543 (1988)\\
$^6$K. Kakurai, M. Steiner, R. Pynn and J. K.  Kjems, J. Phys.: Condens. 

Matter 3, 715 (1991)\\
$^7$A. Zheludev,  J. M. Tranquada, T. Vogt and D. J. Buttrey, Phys. Rev. B 

54, 6437 (1996);
Phys. Rev. B 54, 7210 (1996)\\
$^8$S. Raymond, T. Yokoo, A. Zheludev, S. E. Nagler, A. Wildes and J. 

Akimitsu, Cond-mat/9811040\\
$^9$A. Kl\"umper, A. Schadschneider and J. Zittartz, J.phys. A 24, L955 (1991);

 Z. Phys.B 87, 281 (1992);
Europhys. Lett. 24, 293 (1993)\\
$^{10}$H. Niggemann and J. Zittartz, Z.Phys. B 101, 289 (1996)\\
$^{11}$K. Totsuka and M. Suzuki, J.Phys.: Condens. Matter 7, 1639 (1995)\\
$^{12}$A. Zheludev, Cond-mat/9707167 and references therein.\\
\newpage
\section*{Figure Captions}

Fig. 1. Model mixed spin system consisting of a square lattice of S = 1/2 

spins  (A-spins, solid
circle). The centre of each square plaquette is 

occupied by a S = 1 spin (B-spin, solid
square) .\\
Fig. 2.  A chain of A-spins (upper row) connected to a chain of B-spins (lower

 row).\\

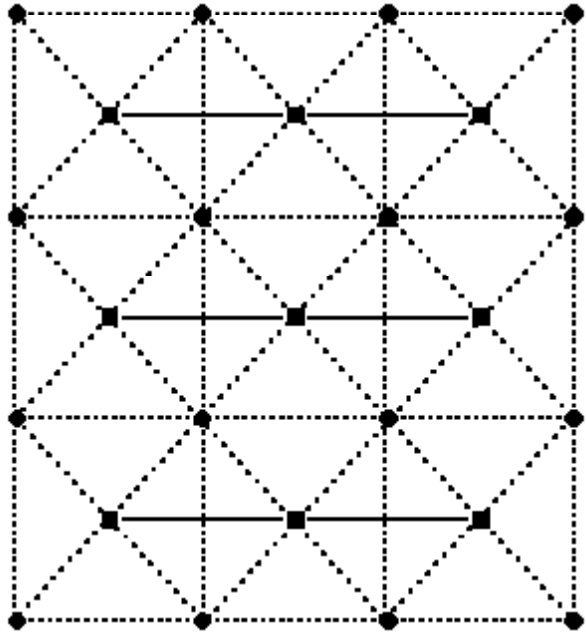
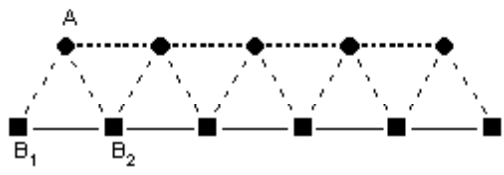
\end{document}